# Quasi Regular Polyhedra and Their Duals with Coxeter Symmetries Represented by Quaternions II

Mehmet Koca[a], Mudhahir Al Ajmi[b]
and
Saleh Al- Shidhani[c]

Department of Physics, College of Science, Sultan Qaboos University
P.O. Box 36, Al-Khoud, 123 Muscat, Sultanate of Oman

**Abstract**
In this paper we construct the quasi regular polyhedra and their duals which are the generalizations of the Archimedean and Catalan solids respectively. This work is an extension of two previous papers of ours which were based on the Archimedean and Catalan solids obtained as the orbits of the Coxeter groups $W(A_3)$, $W(B_3)$ and $W(H_3)$. When these groups act on an arbitrary vector in 3D Euclidean space they generate the orbits corresponding to the quasi regular polyhedra. Special choices of the vectors lead to the platonic and Archimedean solids. In general, the faces of the quasi regular polyhedra consist of the equilateral triangles, squares, regular pentagons as well as rectangles, isogonal hexagons, isogonal octagons, and isogonal decagons depending on the choice of the Coxeter groups of interest. We follow the quaternionic representation of the group elements of the Coxeter groups which necessarily leads to the quaternionic representation of the vertices. We note the fact that the $C_{60}$ molecule can best be represented by a truncated icosahedron where the hexagonal faces are not regular, rather, they are isogonal hexagons where single bonds and double bonds of the carbon atoms are represented by the alternating edge lengths of isogonal hexagons.

[a] electronic-mail: kocam@squ.edu.om
[b] electronic-mail: mudhahir@squ.edu.om
[c] electronic-mail: shidhani7@squ.edu.om

## I. Introduction

The nature seems to follow the symmetries derived from the Coxeter diagrams. The molecular structures [1], viral symmetries [2], crystallographic and quasi crystallographic materials [3] all share the same feature, namely, their symmetries and the vertices of the polyhedra can be represented respectively by the Coxeter groups [4] and their orbits. There exits a one to one correspondence between the quaternionic discrete groups and the 2D, 3D, and 4D Coxeter groups [5]. In two previous papers we have constructed the vertices of the Platonic –Archimedean solids [6] and the dual solids of the Archimedean solids, the Catalan solids [7], using the quaternionic representations of the 3D Coxeter groups. In this paper we use the same technique of references [6-7] to construct the vertices of the quasi regular polyhedra and their dual solids. We call the polyhedra, quasi regular, if they are derived from the Coxeter diagrams and composed of regular as well as isogonal polygons. This excludes the Platonic solids, Archimedean solids and their duals the Catalan solids. In the reference [8] hereafter called the paper I the isogonal polygons consisting of two alternating edge lengths with equal interior angles are constructed. The duals of the isogonal polygons are the isotoxal polygons which have equal edge lengths with alternating two different interior angles. The quasi regular polyhedra that we will be constructing may consist of rectangles, isogonal hexagons, isogonal octagons and isogonal decagons in addition to the equilateral triangles, squares and the regular pentagons.

We are motivated by the fact that the $C_{60}$ molecule has two bond lengths, the single bond, $C-C \approx 1.455 \overset{0}{\text{A}}$ and the double bond, $C=C \approx 1.391 \overset{0}{\text{A}}$ [9] implying that the hexagons of the soccer-ball model are not regular rather they are isogonal hexagons with two alternating edge lengths having $120^0$ interior angles.

We organize the paper as follows. In Section II we introduce the basic facts about the Coxeter diagrams $A_3$, $B_3$ and $H_3$ representing respectively the tetrahedral, octahedral and icosahedral symmetries. For each diagram we define a "highest" weight vector expressed as a linear combination of imaginary quaternionic units. Action of the group elements of the Coxeter groups $W(A_3)$, $W(B_3)$ and $W(H_3)$ on a general "highest" weight vector generate respectively the orbits of sizes 24, 48 and 120. Certain choices of the parameters of the "highest" weight vector lead to the Platonic solids, Archimedean solids as well as the quasi regular polyhedra. Therefore in Section III we discuss the orbits of $W(A_3)$ representing the quasi regular polyhedra and their dual solids possessing the tetrahedral symmetry. The Section IV deals with the orbits of the octahedral group $W(B_3)$ involving quasi regular polyhedra and their dual solids. The orbits of the icosahedral group have been discussed in Section V describing the icosahedral polyhedra and their dual structures. Finally in Section VI we present a brief discussion on our method as to how apply our results to the natural phenomena.

## II. Construction of the Coxeter groups $W(A_3)$, $W(B_3)$, and $W(H_3)$ in terms of quaternions.

Let $q = q_0 + q_i e_i$, ($i = 1,2,3$) be a real unit quaternion with its conjugate defined by $\bar{q} = q_0 - q_i e_i$ and the norm $q\bar{q} = \bar{q}q = 1$. The quaternionic imaginary units satisfy the relations

$$e_i e_j = -\delta_{ij} + \varepsilon_{ijk} e_k, \ (i,j,k = 1,2,3) \tag{1}$$

where $\delta_{ij}$ and $\varepsilon_{ijk}$ are the Kronecker and Levi-Civita symbols and summation over the repeated indices is implicit. The unit quaternions form a group isomorphic to the unitary group $SU(2)$. With the definition of the scalar product

$$(p,q) = \frac{1}{2}(\bar{p}q + \bar{q}p) = \frac{1}{2}(p\bar{q} + q\bar{p}), \tag{2}$$

quaternions generate the four-dimensional Euclidean space.

The Coxeter diagram $A_3$ with its quaternionic roots is shown in **Fig.1**.

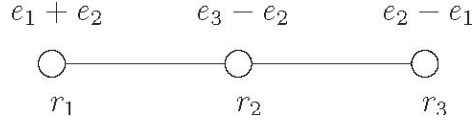

**FIG.1. The Coxeter diagram $A_3$ with quaternionic simple roots.**

The Cartan matrix of the Coxeter diagram $A_3$ and its inverse matrix are given respectively by

$$C = \begin{bmatrix} 2 & -1 & 0 \\ -1 & 2 & -1 \\ 0 & -1 & 2 \end{bmatrix}, \quad C^{-1} = \frac{1}{4}\begin{bmatrix} 3 & 2 & 1 \\ 2 & 4 & 2 \\ 1 & 2 & 3 \end{bmatrix}. \tag{3}$$

The simple roots $\alpha_i$ and their duals $\omega_i$ satisfy the scalar product relations [10]

$$(\alpha_i, \alpha_j) = C_{ij}, \ (\omega_i, \omega_j) = (C^{-1})_{ij}, \ \text{and} \ (\alpha_i, \omega_j) = \delta_{ij}, \ i,j = 1,2,3. \tag{4}$$

We note also that they can be written as linear combinations of each other:

$$\alpha_i = C_{ij}\omega_j, \quad \omega_i = (C^{-1})_{ij}\alpha_j. \tag{5}$$

Let $\alpha$ be an arbitrary simple root given in terms of quaternions. Then the reflection of an arbitrary vector $\Lambda$ with respect to the plane orthogonal to the simple root $\alpha$ is given by [11]

$$r\Lambda = -\frac{\alpha}{\sqrt{2}}\bar{\Lambda}\frac{\alpha}{\sqrt{2}} \equiv [\frac{\alpha}{\sqrt{2}}, -\frac{\alpha}{\sqrt{2}}]^*\Lambda. \tag{6}$$

Therefore the generators of the Coxeter group $W(A_3) \approx T_d \approx S_4$ are given by

$$r_1 = [\frac{1}{\sqrt{2}}(e_1+e_2), -\frac{1}{\sqrt{2}}(e_1+e_2)]^*,$$
$$r_2 = [\frac{1}{\sqrt{2}}(e_3-e_2), -\frac{1}{\sqrt{2}}(e_3-e_2)]^*, \qquad (7)$$
$$r_3 = [\frac{1}{\sqrt{2}}(e_2-e_1), -\frac{1}{\sqrt{2}}(e_2-e_1)]^*.$$

The group elements of the Coxeter group which is isomorphic to the tetrahedral group of order 24 can be written compactly by the set

$$W(A_3) \approx T_d \approx S_4 = \{[p,\overline{p}]\oplus[t,\overline{t}]^*\}, \ p \in T, \ t \in T'. \qquad (8)$$

Here $T$ and $T'$ represent respectively the binary tetrahedral group of order 24 and the coset representative $T' = \frac{O}{T}$ where $O$ is the binary octahedral group of quaternions of order 48 [6]. The Coxeter diagram $B_3$ leading to the octahedral group $W(B_3) \approx O_h$ is shown in **Fig. 2**.

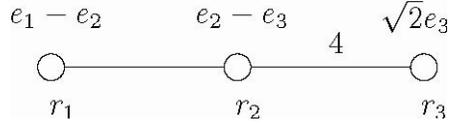

**FIG. 2. The Coxeter diagram $B_3$ with quaternionic simple roots.**

The Cartan matrix of the Coxeter diagram $B_3$ and its inverse matrix are given by

$$C = \begin{bmatrix} 2 & -1 & 0 \\ -1 & 2 & -\sqrt{2} \\ 0 & -\sqrt{2} & 2 \end{bmatrix}, \qquad C^{-1} = \begin{bmatrix} 1 & 1 & \frac{1}{\sqrt{2}} \\ 1 & 2 & \sqrt{2} \\ \frac{1}{\sqrt{2}} & \sqrt{2} & \frac{3}{2} \end{bmatrix}. \qquad (9)$$

The generators,
$$r_1 = [\frac{1}{\sqrt{2}}(e_1-e_2), -\frac{1}{\sqrt{2}}(e_1-e_2)]^*,$$
$$r_2 = [\frac{1}{\sqrt{2}}(e_2-e_3), -\frac{1}{\sqrt{2}}(e_2-e_3)]^*, \ r_3 = [e_3,-e_3]^* \qquad (10)$$
generate the octahedral group which can be written as

$$W(B_3) \approx Aut(A_3) \approx S_4 \times C_2 = \{[p,\overline{p}] \oplus [p,\overline{p}]^* \oplus [t,\overline{t}] \oplus [t,\overline{t}]^*\},$$
$$p \in T, t \in T'. \tag{11}$$

Note that we have three maximal subgroups of the octahedral group $W(B_3)$, namely, the tetrahedral group $W(A_3)$, the chiral octahedral group consisting of the elements of the proper rotations $\frac{W(B_3)}{C_2} = \{[p,\overline{p}] \oplus [t,\overline{t}]\}$, and the pyritohedral group consisting of the elements $T_h \approx A_4 \times C_2 = \{[p,\overline{p}] \oplus [p,\overline{p}]^*\}$. The pyritohedral symmetry represents the symmetry of the pyritohedron which is an irregular dodecahedron with irregular pentagonal faces which occurs in pyrites.

The Coxeter diagram $H_3$ leading to the icosahedral group is shown in **Fig. 3**. There doesn't exist any Lie algebra corresponding to the Coxeter diagram $H_3$.

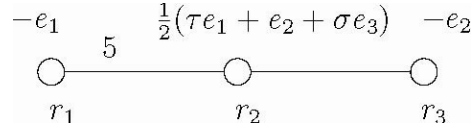

**FIG. 3. The Coxeter diagram of $H_3$ with quaternionic simple roots.**

The Cartan matrix of the diagram $H_3$ and its inverse are given as follows:

$$C = \begin{bmatrix} 2 & -\tau & 0 \\ -\tau & 2 & -1 \\ 0 & -1 & 2 \end{bmatrix}, \quad C^{-1} = \frac{1}{2}\begin{bmatrix} 3\tau^2 & 2\tau^3 & \tau^3 \\ 2\tau^3 & 4\tau^2 & 2\tau^2 \\ \tau^3 & 2\tau^2 & \tau+2 \end{bmatrix}. \tag{12}$$

The generators,
$$r_1 = [e_1, -e_1]^*, r_2 = [\frac{1}{2}(\tau e_1 + e_2 + \sigma e_3), -\frac{1}{2}(\tau e_1 + e_2 + \sigma e_3)]^*, r_3 = [e_2, -e_2]^* \tag{13}$$
generate the icosahedral group
$$I_h \approx W(H_3) = \{[p,\overline{p}] \oplus [p,\overline{p}]^*\} \approx A_5 \times C_2, \quad (p,\overline{p} \in I). \tag{14}$$

Here $I$ is the set of 120 quaternionic elements of the binary icosahedral group [11]. The chiral icosahedral group is represented by the set $W(H_3)/C_2 \approx A_5 = \{[p,\overline{p}]\}$ which is isomorphic to the even permutations of five letters. Note also that the pyritohedral group is a maximal subgroup of the Coxeter group $W(H_3)$.

The general vector in the dual space is represented by the vector $\Lambda = a_1\omega_1 + a_2\omega_2 + a_3\omega_3 \equiv (a_1a_2a_3)$. We will use the notation $O(\Lambda) = W(G)\Lambda = O(a_1a_2a_3)$ for the orbit of the Coxeter group generated from the vector $\Lambda$. We follow the Dynkin notation to represent an arbitrary vector $\Lambda = (a_1a_2a_3)$ in the dual space and drop the basis vectors $\omega_i$, i=1,2,3. In the Lie algebraic representation theory the components $(a_1a_2a_3)$ of the vector $\Lambda$ are called the Dynkin indices [12] which are non-negative integers if it represents the highest weight vector. Here we are not restricted to the integer values of the Dynkin indices. They can be any real number. When the components of the vector in the dual space are not integers we will separate them by commas otherwise no commas will be used. For an arbitrary Coxeter diagram of rank 3 we define the fundamental orbits as

$$O(\omega_1) = O(100),\ O(\omega_2) = O(010),\ \text{and}\ O(\omega_3) = O(001). \tag{15}$$

Any linear combination of the basis vectors $\omega_i$ with real numbers will lead to quasi regular polyhedra, in general, under the action of the Coxeter group. Nevertheless, we will often give examples of orbits consisting of the vectors $(a_1a_2a_3)$ with integer components. In the next three sections we discuss the constructions of the quasi regular polyhedra and their dual solids.

### III. Quasi regular polyhedra with the tetrahedral symmetry $W(A_3)$

We have seen in the paper [6] that the fundamental orbits $O(100)$, $O(010)$, and $O(001)$ represent the tetrahedron, octahedron and dual tetrahedron respectively. Here we have $O(001) = -O(100)$. They play an essential role in the construction of the duals of the Archimedean solids as well as the duals of the quasi regular polyhedra. We note, before proceeding further, that the Coxeter diagram $A_3$ has a Dynkin diagram symmetry which enlarges the group structure to the octahedral symmetry. Therefore the polyhedra obtained from the orbits $O(a_1a_2a_1)$, $a_1 \neq a_2$ possessing the diagram symmetry will be discussed at the end of this section. The non-trivial quasi regular polyhedra with the tetrahedral symmetry are of the types $O(a_1a_20)$, $O(a_10a_3)$, and $O(a_1a_2a_3)$ with $a_1 \neq a_2 \neq a_3$ in general. If the Dynkin indices are equal they are called the Archimedean solids possessing the tetrahedral symmetry. Actually the orbits $O(110)$ and $O(011)$ both represent the truncated tetrahedra. The orbits $O(101)$ and $O(111)$ are respectively the cuboctahedron and the truncated octahedron which have larger octahedral symmetries. Therefore the quasi regular polyhedra invariant only under the tetrahedral symmetry include the orbits
$O(a_1a_20),\ a_1 \neq a_2;\ O(a_10a_3),\ a_1 \neq a_3;\ O(a_1a_1a_3),\ a_1 \neq a_3;$
$O(a_1a_2a_3),\ a_1 \neq a_2 \neq a_3$. (16)

The general orbit $O(a_1a_2a_3)$ which consists of 24 elements is given as follows [6]:

$$\pm \alpha e_1 \pm \beta e_2 \pm \gamma e_3; \pm \beta e_1 \pm \gamma e_2 \pm \alpha e_3; \pm \gamma e_1 \pm \alpha e_2 \pm \beta e_3;$$
$$\pm \alpha e_1 \pm \gamma e_2 \pm \beta e_3; \pm \gamma e_1 \pm \beta e_2 \pm \alpha e_3; \pm \beta e_1 \pm \alpha e_2 \pm \gamma e_3 \qquad (17)$$

even number of minus sign $(-)$ with

$$\alpha = \frac{1}{2}(a_1 - a_3), \beta = \frac{1}{2}(a_1 + a_3), \gamma = \frac{1}{2}(a_1 + 2a_2 + a_3). \qquad (18)$$

Now we consider the polyhedra of (16) and their duals in turn.

**A.   The polyhedron $O(a_1a_20)$, $a_1 \neq a_2$ and its dual solid**

The orbit $O(a_1a_20)$, $a_1 \neq a_2$ is described by the set of 12 vertices since $\alpha = \beta \neq \gamma$. There is a simple way of counting the vertices, edges and the faces of a given polyhedra represented by the Coxeter diagram. The vector $\Lambda = a_1\omega_1 + a_2\omega_2 = (a_1a_20)$ is invariant under the subgroup $C_2$ generated by $r_3$ of the Coxeter group $W(A_3)$. The index of the group $C_2$ in the group $W(A_3)$, $\frac{|W(A_3)|}{|C_2|} = 12$ represents the number of vertices of the quasi regular polyhedron $O(a_1a_20)$. It has 8=4+4 faces (4 triangular faces = $\frac{|W(A_3)|}{|D_3|}$ and 4 isogonal hexagonal faces = $\frac{|W(A_3)|}{|D_3|}$). The edges sharing the vertex $(a_1a_20)$ are represented by the lines parallel to the simple roots $\pm\alpha_1$ and $\pm\alpha_2$. The simple root $\pm\alpha_1$ is invariant under the group $C_2 \times C_2 = \{r_1, r_3\}$ so that the index is $\frac{|W(A_3)|}{|C_2 \times C_2|} = 6$. The root $\pm\alpha_2$ is left invariant by the group $C_2 = \{r_2\}$ so that the index is 12. Therefore the polyhedron $O(a_1a_20)$ has 18 edges. The Euler formula $\chi = V - E + F = 2$ is satisfied. The polyhedron $O(a_1a_20)$ is a truncated tetrahedron such that the faces obtained by the truncation consist of the isogonal hexagons with the alternating edge lengths $\sqrt{2}a_1$ and $\sqrt{2}a_2$ and the equilateral triangles of sides $\sqrt{2}a_2$. The quasi regular truncated tetrahedron is shown in **Fig. 4(a)** for the values $a_1 = 1$ and $a_2 = 2$. At the vertex $(a_1a_20)$ we have two isogonal hexagons and one equilateral triangle.

Now we discuss how to construct the dual of this quasi regular truncated solid. The isogonal hexagon $O(a_1a_2)$ is obtained by the action of the dihedral group $D_3$ generated by $r_1$ and $r_2$ on the vector $(a_1a_2)$ leaving the vector $\omega_3$ invariant [8]. This indicates that the vector $\omega_3$ is orthogonal to the plane of the isogonal hexagon and can be taken as the center of the hexagon up to a scale factor.

Similarly the triangle $O(a_2 0)$ can be obtained by the dihedral group generated by $r_2$ and $r_3$ whose center can be represented by the vector $\omega_1$ up to a scale factor. Then it is clear that the centers of two isogonal hexagons joining to the vertex $\Lambda = (a_1 a_2 0)$ are the vectors $\omega_3$ and $r_3 \omega_3 = \omega_3 - \alpha_3$. Then these three vectors $\omega_1$, $\omega_3$, and $\omega_3 - \alpha_3$ form an isosceles triangle. To obtain a particular face of the dual polyhedron of the quasi regular polyhedron $O(a_1 a_2 0)$ the isosceles triangle must be orthogonal to the vertex $\Lambda = (a_1 a_2 0)$. This can be achieved by arranging the relative lengths of the vectors $\omega_1$ and $\omega_3$. Since $\omega_3 - r_3 \omega_3 = \alpha_3$ is orthogonal to the vector $\Lambda = (a_1 a_2 0)$ it is sufficient to compute the relative magnitude of $\lambda \omega_1$ from the relation

$$(\lambda \omega_1 - \omega_3).(a_1 \omega_1 + a_2 \omega_2) = 0 \tag{19}$$

which leads to the value

$$\lambda = \frac{a_1 + 2a_2}{3a_1 + 2a_2}. \tag{20}$$

When $a_1 = a_2$ the polyhedron is a truncated tetrahedron with regular hexagonal and equilateral triangular faces and its dual is the Catalan solid with $\lambda = 0.6$ called triakis tetrahedron [7]. For an arbitrary $a_1$ and $a_2$ the dual of the quasi regular polyhedron still has the faces of isosceles triangles but different from the faces of the Catalan solid. For example, the dual solid of the quasi regular tetrahedron $O(120)$ consists of 4+4=8 vertices obtained from two fundamental orbits $\lambda O(100)$ and $O(001)$ where $\lambda = \frac{5}{7}$. The vertices of the dual solid lie on two concentric spheres with radii $\frac{5\sqrt{3}}{14}$ and $\frac{\sqrt{3}}{2}$. It is shown in the **Fig. 4(b)**.

**B. The polyhedron $O(a_1 0 a_3)$ and its dual solid**

With the similar argument above we can show that this is a quasi regular polyhedron with 12 vertices, 24 edges, 4 triangular faces of edge length $\sqrt{2} a_1$, 4 triangular faces of edge length $\sqrt{2} a_3$, and 6 rectangular faces with edges $\sqrt{2} a_1$ and $\sqrt{2} a_3$. When we have $a_1 = a_3$ we obtain a *cuboctahedron*. However for a general vector with $a_1 \neq a_3$ one obtains a quasi regular polyhedron with rectangular and triangular faces. As an example we have displayed the solid of the orbit $O(102)$ in **Fig. 5(a)**.

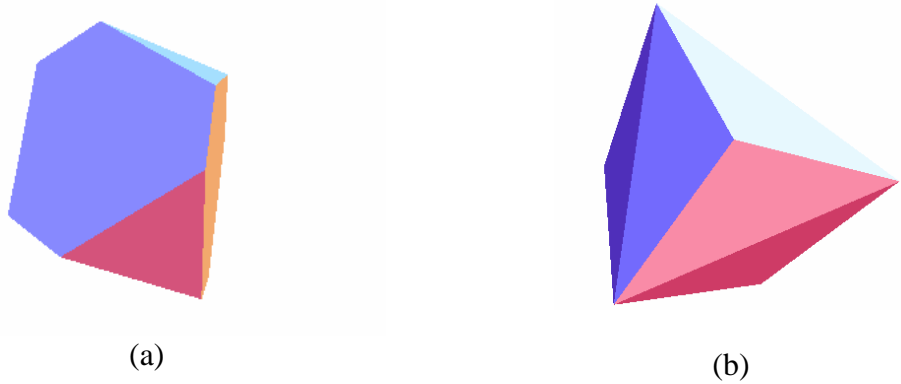

(a)                  (b)

**FIG. 4. Truncated quasi regular tetrahedron (a) and its dual polyhedron (b).**

Sharing the vertex $\Lambda = (a_1 0 a_3)$ we have two rectangles and two triangles. Centers of the faces, up to some scale factors, are represented by the vectors as follows:

Centers of the rectangles           : $\omega_2$ and $r_2\omega_2 = \omega_2 - \alpha_2$,

Center of the triangle of edge $\sqrt{2}a_1$ : $\omega_3$,

Center of the triangle of edge $\sqrt{2}a_3$ : $\omega_1$.

Now we discuss the dual of this solid. The vertex $\Lambda = (a_1 0 a_3)$ is left invariant by the group $C_2$ generated by $r_2$. The face orthogonal to this vector is a kite with the same symmetry $C_2$. We can determine the relative lengths of the vectors $\lambda\omega_1$ and $\eta\omega_3$ from the equations

$(\lambda\omega_1 - \omega_2).(a_1\omega_1 + a_3\omega_3) = 0$,     $(\eta\omega_3 - \omega_2).(a_1\omega_1 + a_3\omega_3) = 0$,

$$\lambda = \frac{2a_1 + 2a_3}{3a_1 + a_3}, \qquad \eta = \frac{2a_1 + 2a_3}{a_1 + 3a_3}. \tag{21}$$

The dual solid with 12 kite faces, 24 edges has 14=4+4+6 vertices consisting of the orbits $\lambda O(100)$, $\eta O(001)$, and $O(010)$. The vertices of the dual solid of the polyhedron $O(102)$ lie on three concentric spheres with relative radii are given by $\dfrac{3\sqrt{3}}{7} : 1 : \dfrac{3\sqrt{3}}{5}$. The dual of the solid $O(102)$ is depicted in **Fig. 5(b)**.

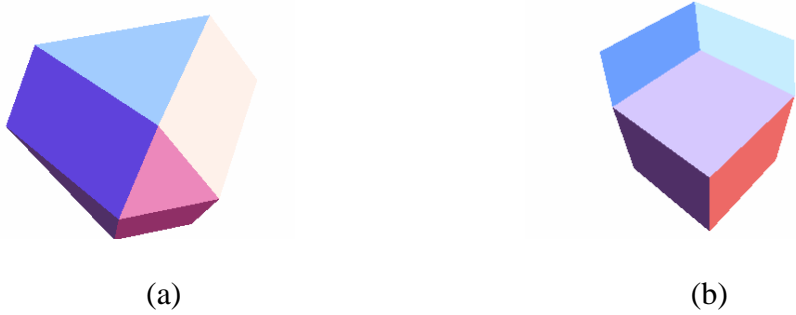

(a)                           (b)

**FIG.5. The quasi regular polyhedron of the orbit $O(102)$ (a) and its dual solid (b).**

### C. The polyhedron $O(a_1 a_2 a_3)$ and its dual solid

Let us consider the most general case where $a_1 \neq a_2 \neq a_3$. This solid has 24 vertices, 36 edges, and 14 faces (4 isogonal hexagons of edge lengths $\sqrt{2}a_1$ and $\sqrt{2}a_2$, 4 isogonal hexagons of edge lengths $\sqrt{2}a_2$ and $\sqrt{2}a_3$, 6 rectangles of edges $\sqrt{2}a_1$ and $\sqrt{2}a_3$). The three faces share the same vertex $\Lambda=(a_1 a_2 a_3)$. The solid represented by the orbit $O(123)$ is depicted as a model example in **Fig. 6(a)**. The symmetry leaving the vector $\Lambda=(a_1 a_2 a_3)$ invariant is trivial so that the face of the dual solid is a scalene triangle. The vertices of the face of the dual solid orthogonal to the vertex $\Lambda$ is determined from the relations

$$(\lambda \omega_1 - \omega_3).\Lambda = 0, \text{ and } (\eta \omega_2 - \omega_3).\Lambda = 0, \tag{22}$$

leading to

$$\lambda = \frac{a_1 + 2a_2 + 3a_3}{3a_1 + 2a_2 + a_3}, \quad \eta = \frac{a_1 + 2a_2 + 3a_3}{2a_1 + 4a_2 + 2a_3}. \tag{23}$$

The vertices of the dual solid $O(123)$ consist of the fundamental orbits

$$1.4 O(100), \ \frac{7}{8} O(010), \text{ and } O(001), \tag{24}$$

which can be determined using (17) and (18). The dual solid of the polyhedron $O(123)$ is shown in **Fig. 6(b).**

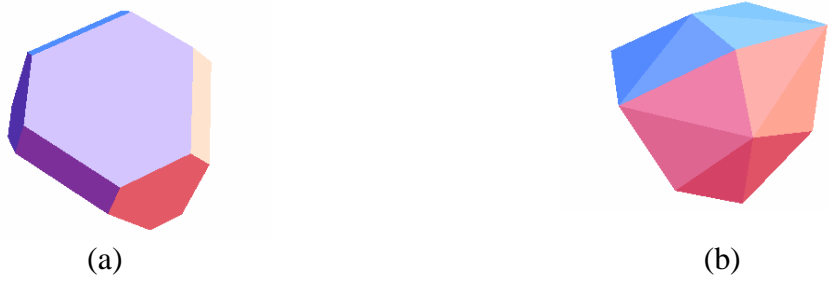

**FIG.6. The polyhedron** $O(123)$ **(a) and its dual solid (b).**

There are two special cases worth mentioning. If $a_1 = a_2$ then one of the isogonal hexagon turns out to be a regular hexagon in the polyhedron $O(a_1 a_1 a_3)$, $a_1 \neq a_3$. Its dual still consists of faces which are scalene triangles. Another special case is that where the indices satisfy $a_1 = a_3 \neq a_2$. This polyhedron has a larger octahedral symmetry which will be discussed in the next section.

### D. The polyhedron $O(a_1 a_2 a_1)$ and its dual

Although this solid has an additional symmetry as we mentioned and will be discussed in the next section we would also give its structure with the tetrahedral symmetry. This solid has 24 vertices, 36 edges, and 14 faces (8 isogonal hexagons of edge lengths $\sqrt{2}a_1$ and $\sqrt{2}a_2$, 6 squares of edge $\sqrt{2}a_1$). The three faces join each other at the vertex $\Lambda = (a_1 a_2 a_1)$. The solid represented by the orbit $O(121)$ is depicted in **Fig. 7(a)**.

The centers of the isogonal hexagons joining to the vertex $\Lambda = (a_1 a_2 a_1)$ can be taken as $\omega_1$ and $\omega_3$ and the center of the square is $\omega_2$. The faces of the dual of the solid are isosceles triangles and the vertices of the triangle which is orthogonal to the vertex $\Lambda = (a_1 a_2 a_1)$ are given by the vectors $\omega_1$, $\lambda\omega_2$, and $\omega_3$ where $\lambda = \dfrac{2a_1 + a_2}{2(a_1 + a_2)}$. The vertices of the dual of the polyhedron $O(121)$ are given by the fundamental orbits $O(100)$, $\dfrac{2}{3}O(010)$ and $O(001)$. It is depicted in **Fig. 7(b)**.

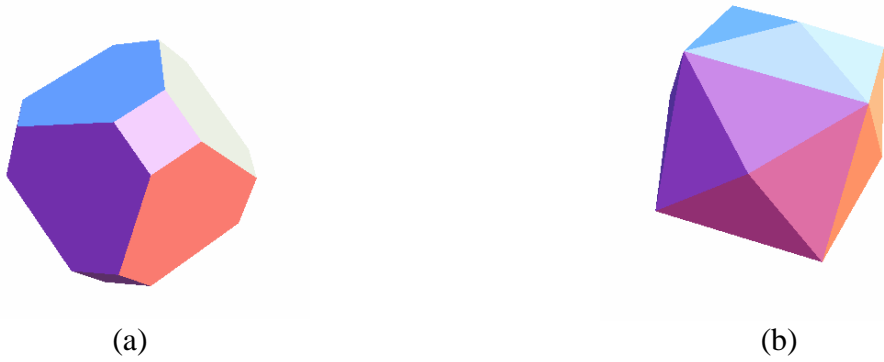

**FIG.7. The polyhedron** $O(121)$ **(a) and its dual (b).**

There is a very interesting quasi regular polyhedron whose faces consist of the squares with edge length of the golden ratio $\tau=\dfrac{1+\sqrt{5}}{2}$ and the isogonal hexagons with edges equal to 1 and $\tau$. The vertices follow from the orbit $W(A_2)(a_1 a_2 a_3)$ where $a_1 = a_3 = \tau, a_2 = 1$. The substitution of these values in (18) and (17) will lead to two set of vectors up to a scale factor

$$\pm e_1 \pm \tau e_2, \quad \pm e_2 \pm \tau e_3, \quad \pm e_3 \pm \tau e_1; \qquad (25a)$$
$$\pm \tau e_1 \pm e_2, \quad \pm \tau e_2 \pm e_3, \quad \pm \tau e_3 \pm e_1. \qquad (25b)$$

The 12 vertices of (25a) and the second set of vectors of (25b) represent two different icosahedra. Actually the first set is an orbit obtained by acting the subgroup $(W(A_2)/C_2)(\tau,1,\tau)$. The group $(W(A_2)/C_2) \approx A_4$ contains only the rotational elements of the group. Therefore the orbit is a chiral icosahedron [13]. The vectors of (25b) can be obtained by acting the group $(W(A_2)/C_2)$ on the vector $r_1(\tau,1,\tau)$. The vector $r_1(\tau,1,\tau)$ is the mirror reflection of $(\tau,1,\tau)$ under the generator $r_1$. Therefore the set of vectors of (25a-b) consists of two orbits which are mirror symmetry of two chiral icosahedrons. The quasi regular solid whose vertices are given in (25a-b) and its dual are shown in **Fig. 8**.

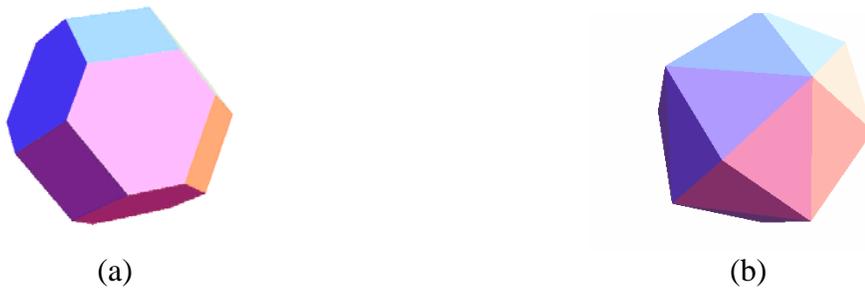

**FIG.8. The polyhedron** $O(\tau,1,\tau)$ **(a) and its dual solid (b).**

## IV. Quasi regular polyhedra with the octahedral symmetry $W(B_3)$

The general orbit $O(a_1 a_2 a_3)$ consists of 48 vectors given as follows [6]:

$$\pm \alpha e_1 \pm \beta e_2 \pm \gamma e_3; \pm \beta e_1 \pm \gamma e_2 \pm \alpha e_3; \pm \gamma e_1 \pm \alpha e_2 \pm \beta e_3;$$
$$\pm \alpha e_1 \pm \gamma e_2 \pm \beta e_3; \pm \gamma e_1 \pm \beta e_2 \pm \alpha e_3; \pm \beta e_1 \pm \alpha e_2 \pm \gamma e_3 \quad (26)$$

with

$$\alpha = (a_1 + a_2 + \frac{a_3}{\sqrt{2}}), \beta = (a_2 + \frac{a_3}{\sqrt{2}}), \gamma = \frac{a_3}{\sqrt{2}}. \quad (27)$$

Here the fundamental orbits $O(100), O(010)$, and $O(001)$ represent respectively the octahedron, cuboctahedron, and the cube. Since the Archimedean solids $O(110), O(101), O(011)$, and $O(111)$ and their duals, the Catalan solids, are respectively discussed in the references [6] and [7] we will here consider only the quasi regular polyhedra and their dual solids possessing the octahedral symmetry. Now let us deal with them in turn.

### A. The orbit $O(a_1 a_2 0)$, $a_1 \neq a_2$ and its dual

This is the solid discussed in the section III.D. Now using the octahedral symmetry we determine again its vertices, edges, and faces. Then it turns out to be a quasi regular truncated octahedron with 24 vertices, 36 edges and 14 faces (8 isogonal hexagons with the alternating edge lengths $\sqrt{2}a_1$ and $\sqrt{2}a_2$ and 6 squares with edges $\sqrt{2}a_2$). The vertices of the polyhedron $O(210)$ is obtained by substituting $a_1 = 2, a_2 = 1$ in (27) and then in (26). It has exactly the same set of vertices of the orbit $O(121)$ of $W(A_3)$. It is plotted in **Fig. 9(a).**

The faces of the dual of the polyhedron $O(a_1 a_2 0)$ are isosceles triangles and the vertices of a typical face are given by the vectors $\lambda \omega_1$, $\omega_3$, and $r_3 \omega_3 = \omega_3 - \alpha_3$ where $\lambda$ is obtained from the relation $(\lambda \omega_1 - \omega_3).(a_1 \omega_1 + a_2 \omega_2) = 0$:

$$\lambda = \frac{a_1 + 2a_2}{\sqrt{2}(a_1 + a_2)} \quad (28)$$

The 14 vertices of the dual solid of the quasi regular truncated octahedron $O(120)$ are the union of two fundamental orbits $\frac{5}{3\sqrt{2}}O(100)$ and $O(001)$. The 6 vertices of the dual solid is on a sphere with radius $\frac{5}{3\sqrt{2}} \approx 1.179$ and the other 8 vertices lie on the concentric sphere with radius $\sqrt{\frac{3}{2}} \approx 1.225$. It is depicted in **Fig.9(b)**. If we let $a_1 = a_2 = a$ we obtain the orbit $O(a,a,0)$ which represents a truncated octahedron commonly known as the Wigner-Seitz cell for the body-centered cubic lattice.

The polyhedron in **Fig.8(a)** can be also obtained from the vector $\Lambda = (1, \tau, 0)$ by applying the group $W(B_3)$ on $\Lambda$. To obtain the vertices we substitute $a_1 = 1$, $a_2 = \tau$ and $a_3 = 0$ in (26).

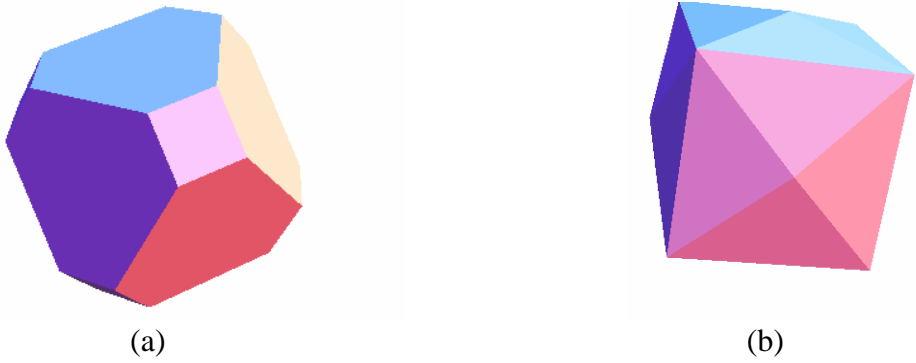

(a)                          (b)

**FIG. 9. The polyhedron $O(210)$ (a) and its dual solid (b)**

**B**. **The polyhedron $O(a_1 0 a_3)$, $a_1 \neq a_3$ and its dual solid**

This is another quasi regular polyhedron with 24 vertices, 48 edges and 26 faces (8 equilateral triangles of the edge length $\sqrt{2}a_1$, 6 squares of the edge length $\sqrt{2}a_3$, and 12 rectangles with the edge lengths $\sqrt{2}a_1$ and $\sqrt{2}a_3$). When the indices are equal, $a_1 = a_3$, then the Archimedean polyhedron is called the *small rhombicuboctahedron* which was discussed in ref. [6]. One equilateral triangle of the edge length $\sqrt{2}a_1$, one square of the edge length $\sqrt{2}a_3$, and two rectangles of edge lengths $\sqrt{2}a_1$ and $\sqrt{2}a_3$ share the vertex $\Lambda = (a_1 0 a_3)$. We construct the polyhedron $O(102)$ whose vertices are obtained by substituting $a_1 = 1$, $a_2 = 0$, $a_3 = 2$ in (27) and then in (26). The polyhedron $O(102)$ is shown in **Fig. 10(a).**

The dual of this polyhedron can be determined as follows. The center of the triangle adjacent to the vertex $\Lambda = (a_1 0 a_3)$ can be represented by the vector $\omega_3$. The center of the square can be taken as $\omega_1$ since the square is left invariant by the generator $r_1$. The centers of two rectangles sharing the vertex $\Lambda = (a_1 0 a_3)$ can be taken as $\omega_2$, and $r_2 \omega_2 = \omega_2 - \alpha_2$. Note that the line joining the centers of the rectangles is orthogonal to the vector $\Lambda = (a_1 0 a_3)$. These four vertices form two isosceles triangles sharing the same edge. When they are on the same plane they represent a kite. We determine the other two vertices of the kite by the requirement that the plane of the kite must be orthogonal to the vertex $\Lambda = (a_1 0 a_3)$ which leads to the equations

$$(\lambda \omega_1 - \omega_2).\Lambda = 0, \text{ and } (\eta \omega_3 - \omega_2).\Lambda = 0. \tag{29}$$

The parameters are determined as follows:

$$\lambda = \frac{\sqrt{2}a_1 + 2a_3}{\sqrt{2}a_1 + a_3}, \ \eta = \frac{2a_1 + 2\sqrt{2}a_3}{\sqrt{2}a_1 + 3a_3}. \tag{30}$$

The vertices of the dual polyhedron are the union of the fundamental weights:

$$\lambda O(100), O(010), \text{ and } \eta O(001). \tag{31}$$

By substituting $a_1 = 1$, $a_2 = 0$ and $a_3 = 2$ in (30) we obtain the set of 26 vertices of the dual polyhedron $\frac{\sqrt{2}+4}{\sqrt{2}+2} O(100), O(010), \text{ and } \frac{2+4\sqrt{2}}{6+\sqrt{2}} O(001)$. These orbits determine three spheres of radii respectively 1.5858, 1.4142, and 1.2648. The dual polyhedron with the above 26 vertices is depicted in **Fig.10(b)**.

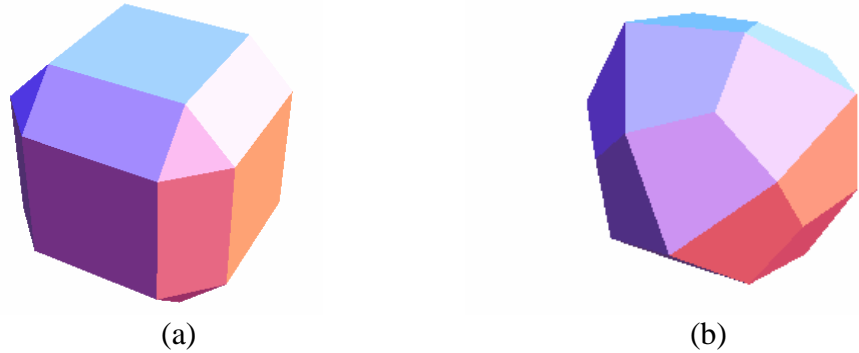

(a)                  (b)

**FIG.10. The regular polyhedron $O(102)$ (a) and its dual solid (b).**

## C. The polyhedron $O(0 a_2 a_3)$, $a_2 \neq a_3$ and its dual solid

If the indices were equal $a_2 = a_3$ then the polyhedron would be a truncated cube with triangular and octagonal faces. For different values of the indices it will represent a quasi regular polyhedron with 24 vertices, 36 edges and 14 faces (8 equilateral triangles with edge length $\sqrt{2}a_2$, 6 isogonal octagons with the alternating edge lengths $\sqrt{2}a_2$ and $\sqrt{2}a_3$). One can obtain the vertices of the polyhedron $O(012)$ from (27) and (26) and it is depicted in **Fig.11(a).**

The center of the triangular face can be taken $\omega_3$ and the centers of the two isogonal octagons at the vertex $\Lambda = (0a_2a_3)$ are the vectors $\omega_1$ and $r_1\omega_1 = \omega_1 - \alpha_1$. Since the line joining the last two vectors is orthogonal to the vector $\Lambda = (0a_2a_3)$ the relative length of the vector $\omega_3$ is determined from $(\lambda\omega_3 - \omega_1).\Lambda = 0$ as

$$\lambda = \frac{2a_2 + \sqrt{2}a_3}{2\sqrt{2}a_2 + 3a_3}. \tag{32}$$

Faces of the dual polyhedron are isosceles triangles. The 14 vertices of the dual solid of the polyhedron $O(012)$ are the union of the fundamental orbits $\frac{1+\sqrt{2}}{3+\sqrt{2}}(001)$ and $(100)$. The dual polyhedron is shown in **Fig.11(b).**

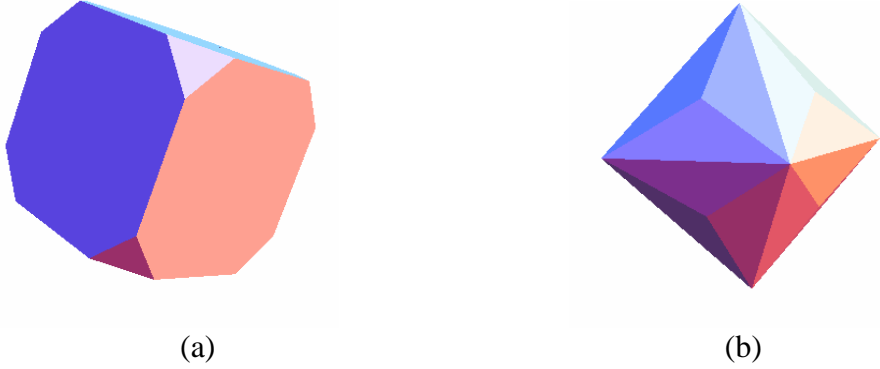

(a)                           (b)

**FIG.11. The quasi regular polyhedron $O(012)$ (a) and its dual solid (b).**

### D. The polyhedron $O(a_1a_2a_3)$ and its dual solid

There are four different cases here. We will discuss first the case where all three indices are different. Then we will consider three special cases where two indices are equal. The polyhedron with all indices different has 48 vertices as the vector $\Lambda = (a_1a_2a_3)$ is left invariant only by the trivial element of the octahedral group. It has 72 edges and 26 faces (8 isogonal hexagons of the edge lengths $\sqrt{2}a_1$ and $\sqrt{2}a_2$, 12 rectangles of the edge lengths $\sqrt{2}a_1$ and $\sqrt{2}a_3$, 6 isogonal octagons of the edge lengths $\sqrt{2}a_2$ and $\sqrt{2}a_3$). These three different faces meet at one vertex.

The dual solid can be obtained as follows. The centers of those three faces meeting at the vertex $\Lambda = (a_1a_2a_3)$ can be represented by the fundamental weights $\omega_1$, $\omega_2$, and $\omega_3$. Of course, the relative lengths of the vectors will be determined from the fact that the vector $\Lambda = (a_1a_2a_3)$ will be orthogonal to the scalene triangle determined by the vectors $\lambda\omega_1$, $\eta\omega_2$, and $\omega_3$

$$(\lambda\omega_1 - \omega_3).\Lambda = 0, \text{ and } (\eta\omega_2 - \omega_3).\Lambda = 0, \tag{33}$$

which leads to

$$\lambda = \frac{\sqrt{2}a_1 + 2\sqrt{2}a_2 + 3a_3}{2a_1 + 2a_2 + \sqrt{2}a_3}, \qquad \eta = \frac{\sqrt{2}a_1 + 2\sqrt{2}a_2 + 3a_3}{2a_1 + 4a_2 + 2\sqrt{2}a_3}. \qquad (34)$$

We give a plot of the polyhedron $O(123)$ and its dual in **Fig.12(a)** and **Fig.12(b)** respectively. The vertices of the polyhedron $O(123)$ are determined from (27) and (26) and the vertices of the dual polyhedron is the union of the fundamental vertices

$$\frac{5\sqrt{2}+9}{6+3\sqrt{2}}O(100), \quad \frac{5\sqrt{2}+9}{10+2\sqrt{2}}O(010), \text{ and } O(001). \qquad (35)$$

They determine three concentric spheres of radii 1.569, 1.772, and 1.225 each having 6, 12, and 8 vertices respectively. The dual polyhedron is face-transitive.

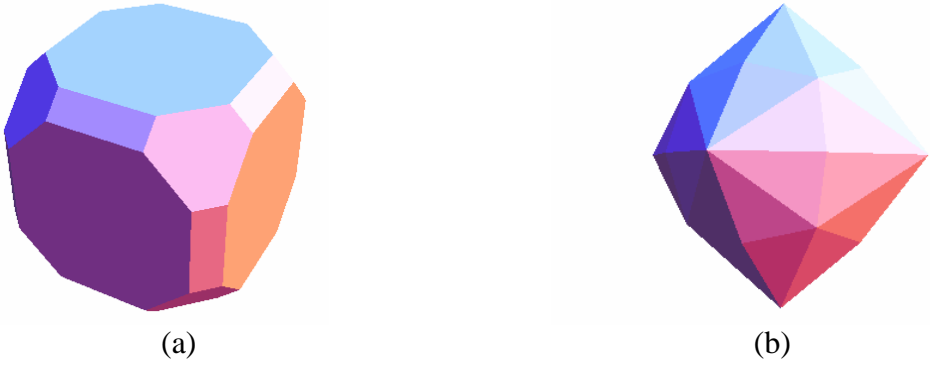

(a)          (b)

**FIG.12.** **The quasi regular polyhedron $O(123)$ (a) and its dual solid (b)**

Now we discuss some special polyhedra obtained from $O(a_1 a_2 a_3)$.

**D1. Polyhedron** $O(a_1 a_1 a_3)$ with $a_1 \neq a_3$

The only difference of this polyhedron from the above one is that the isogonal hexagon in the polyhedron $O(a_1 a_2 a_3)$ now becomes a regular hexagon. The number of vertices, edges and the faces are the same as in the polyhedraon $O(a_1 a_2 a_3)$. The faces of the dual of this polyhedron are also scalene triangles. The values of the parameters $\lambda$ and $\eta$ can be obtained from (34) by letting $a_1 = a_2$.

**D2. Polyhedron** $O(a_1a_2a_1), a_1 \neq a_2$

Here we have the squares of edge length $\sqrt{2}a_1$ and the isogonal hexagons and the isogonal octagons of edge lengths $\sqrt{2}a_1$ and $\sqrt{2}a_2$. The number of vertices, edges and the faces do not change. Faces of the dual polyhedron is again a scalene triangle where the parameters $\lambda$ and $\eta$ can be obtained from (34) by letting $a_1 = a_2$.

**D3. Polyhedron** $O(a_1a_2a_2), a_1 \neq a_2$

The 26 faces of this polyhedron consist of the isogonal hexagons, rectangles and the regular octagons. The dual polyhedron is covered by the 26 scalene triangles whose parameters $\lambda$ and $\eta$ are determined from (34) by letting $a_3 = a_2$.

**V.  Quasi regular polyhedra with the icosahedral symmetry** $W(H_3)$

A general vector for an arbitrary orbit of $W(H_3)$ can be expressed in terms of quaternions as $\Lambda = \sigma(a_1a_2a_3) = \frac{1}{2}[-\sigma a_1 e_1 - \sigma a_3 e_2 + (\tau a_1 + 2a_2 + a_3)e_3]$. This is to set the scale so that the fundamental orbit $\sigma(010)$ represents the root system of $H_3$ in terms of unit quaternions. The general orbit $O[\sigma(a_1a_2a_3)]$ consists of 120 vectors which we do not display them all here because they occupy unnecessary large space. The fundamental orbits $O[\sigma(100)]$, $O[\sigma(010)]$ and $O[\sigma(001)]$ respectively represent a dodecahedron with 20 vertices, an icosidodecahedron with 30 vertices, and an icosahedron with 12 vertices. The Archimedean polyhedra $O[\sigma(110)]$, $O[\sigma(101)]$, $O[\sigma(011)]$ and $O[\sigma(111)]$ represent respectively the truncated dodecahedron, small rhombicosidodecahedron, truncated icosahedron, and the great rhombicosidodecahedron. They are constructed in the reference [6] and their duals, the Catalan solids, are discussed in the reference[7]. Now we discuss each quasi regular polyhedra in turn.

**A.  The polyhedron** $O[\sigma(a_1a_20)]$, $a_1 \neq a_2$ **and its dual solid**

This quasi regular polyhedron has 60 vertices, 90 edges, and 32 faces (20 isogonal decagonal faces of edge lengths $\sqrt{2}a_1$ and $\sqrt{2}a_2$, 12 triangular faces of edges $\sqrt{2}a_2$).

Sharing the vertex $\Lambda = \sigma(a_1a_20)$ we have two isogonal decagons and one triangle and their centers can be represented respectively by the vectors $\omega_3, r_3\omega_3 = \omega_3 - \alpha_3$, and $\lambda \omega_1$ up to an over all scale factor. The factor $\lambda$ is determined as

$$\lambda = \frac{\tau a_1 + 2a_2}{3a_1 + 2\tau a_2}. \qquad (36)$$

For $a_1 = 1$ and $a_2 = 2$ the quasi regular solid and its dual are shown in **Fig. 13**. When $a_1 = a_2 = a$ we get truncated dodecahedron which is an Archimedean solid. The Catalan solid which is the dual of truncated dodecahedron consist of two orbits as above but the scale factor is modified for $a_1 = a_2 = a$.

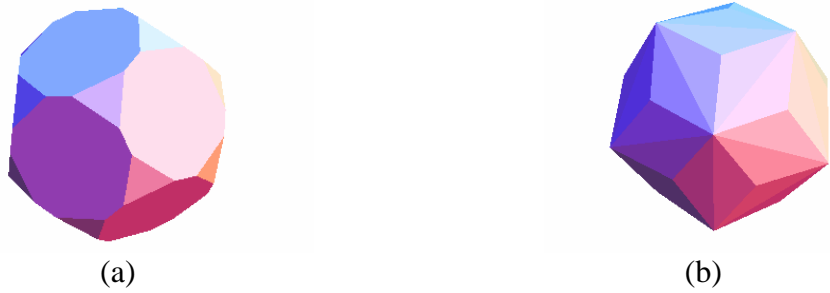

(a)          (b)

**FIG.13. The quasi regular polyhedron $O[\sigma(120)]$ (a) and its dual solid (b)**

**B. The orbit $O[\sigma(a_1 0 a_3)]$, $a_1 \neq a_3$ and its dual**

When $a_1 = a_3 = a$ we obtain the Archimedean solid small rhombicosidodecahedron. The quasi regular polyhedron with $a_1 \neq a_3$ has 60 vertices, 120 edges of lengths $\sqrt{2}a_1$ and $\sqrt{2}a_3$ and 62 faces (12 pentagons of length $\sqrt{2}a_1$, 20 equilateral triangles of edge length $\sqrt{2}a_3$, 30 rectangles of edge lengths $\sqrt{2}a_1$ and $\sqrt{2}a_3$). Joining to the vertex $\sigma(a_1 0 a_3)$ we have one triangle, two rectangles and one pentagon. Their centers can be chosen as $\lambda \omega_1$, $\omega_2$, $r_2 \omega_2$, $\eta \omega_3$ respectively. The scale parameters can be determined as

$$\lambda = \frac{2\tau a_1 + 2a_3}{3a_1 + \tau a_3}, \qquad \eta = \frac{2\tau a_1 + 2a_3}{\tau a_1 + (\sigma + 2)a_3}. \qquad (37)$$

The quasi regular polyhedron and its dual are depicted in **Fig. 14**.

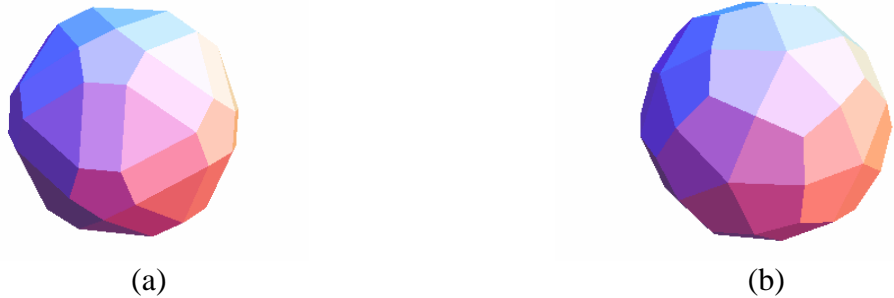

(a)          (b)

**FIG.14. The quasi regular polyhedron $O[\sigma(102)]$ (a) and its dual solid (b).**

## C. The polyhedron $O[\sigma(0a_2a_3)]$, $a_2 \neq a_3$ and its dual

When $a_2 = a_3 = a$ we obtain the Archimedean solid, the truncated icosahedron, with 60 vertices, 90 edges and 32 faces (20 regular hexagons and 12 pentagons). The quasi regular polyhedron for $a_2 \neq a_3$ has also 60 vertices, 90 edges of lengths $\sqrt{2}a_2$ and $\sqrt{2}a_3$ and 32 faces (12 pentagons of length $\sqrt{2}a_2$, 20 isogonal hexagons with edge lengths $\sqrt{2}a_2$ and $\sqrt{2}a_3$). Sharing the vertex $\sigma(0a_2a_3)$ we have two isogonal hexagons and one pentagon. Their centers can be chosen as $\omega_1$, $r_1\omega_1$, $\lambda\omega_3$ respectively and the scale parameter can be determined as

$$\lambda = \frac{\tau(2a_2 + a_3)}{2a_2 + (\sigma + 2)a_3}. \tag{38}$$

The $C_{60}$ molecule is modeled as a truncated icosahedron with regular hexagons. However the experimental values show that the double bond represented by the edge between two hexagons is shorter than the single bond represented by the edge between a hexagon and a pentagon. As we mentioned in Sec.I experimental values of the single and double bond lengths are given respectively as $C-C \approx 1.455 \overset{0}{\text{A}}$ and $C=C \approx 1.391 \overset{0}{\text{A}}$. The bond angles are also measured. They are found to be: $C=C-C$ bond angle=$120^0$; the $C-C-C$ bond angle=$108^0$ [14]. As we pointed out in [8] the interior angle of any isogonal hexagon is $120^0$ and naturally the interior angle of a pentagon is $108^0$. Therefore the actual model of the $C_{60}$ molecule is a truncated icosahedron with isogonal hexagons and pentagons with the ratio of edge lengths $\frac{a_2}{a_3} = \frac{1.455}{1.391} \approx 1.046$. Although this ratio is small it is different from 1.

The $C_{60}$ molecule with the experimental value above and its dual are depicted in **Fig. 15**. When one looks at the figure it is of course difficult to distinguish the polyhedron in **Fig.15(a)** from the regular truncated icosahedron where $a_2 = a_3$ because the ratio $\frac{a_2}{a_3} = \frac{1.455}{1.391} \approx 1.046$ is too small to notice by eye.

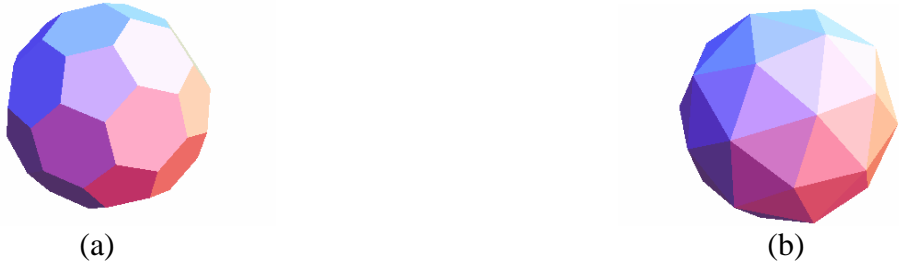

(a)          (b)

**FIG.15. The $C_{60}$ molecule with its experimental bond lengths (a) and its dual solid (b).**

### D.  The orbit $O[\sigma(a_1 a_2 a_3)]$ and its dual

Here also we will treat the orbits in four classes. We will discuss first the case where all three indices are different. Then we will consider three special cases where two indices are equal. The polyhedron with all indices different has 120 vertices as the vector $\Lambda = \sigma(a_1 a_2 a_3)$ is left invariant only by the trivial element of the icosahedral group. It has 180 edges and 62 faces (12 isogonal decagons of edge lengths $\sqrt{2}a_1$ and $\sqrt{2}a_2$, 30 rectangles of edge lengths $\sqrt{2}a_1$ and $\sqrt{2}a_3$, 20 isogonal hexagons of edge lengths $\sqrt{2}a_2$ and $\sqrt{2}a_3$). These three different faces meet at one vertex.

The dual solid can be obtained as follows. The centers of those three faces meeting at the vertex $\Lambda = \sigma(a_1 a_2 a_3)$ can be represented by the fundamental weights $\omega_1$, $\omega_2$, and $\omega_3$. Of course, the relative lengths of the vectors will be determined from the fact that the vector $\Lambda = \sigma(a_1 a_2 a_3)$ will be orthogonal to the scalene triangle determined by the vectors $\lambda \omega_1$, $\eta \omega_2$, and $\omega_3$

$$(\lambda \omega_1 - \omega_3).\Lambda = 0, \text{ and } (\eta \omega_2 - \omega_3).\Lambda = 0, \tag{39}$$

which leads to

$$\lambda = \frac{\tau a_1 + 2a_2 + (2+\sigma)a_3}{3a_1 + 2\tau a_2 + \tau a_3}, \quad \lambda = \frac{\tau a_1 + 2a_2 + (2+\sigma)a_3}{2\tau a_1 + 4a_2 + 2a_3}. \tag{40}$$

With the choice of the vector $\Lambda = \sigma(123)$ the solid represented by the orbit $O[\sigma(123)]$ and its dual are shown in **Fig. 16**.

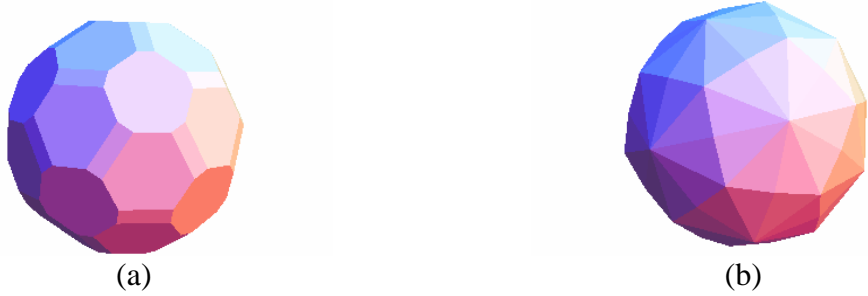

      (a)                           (b)

**FIG.16.**  The quasi regular polyhedron $O[\sigma(123)]$ **(a)** and its dual solid **(b)**.

More symmetric polyhedra can be obtained from $O[\sigma(a_1 a_2 a_3)]$ by choosing some indices equal and will be discussed below.

**D1. The polyhedron** $O[\sigma(a_1a_1a_3)]$ with $a_1 \neq a_3$

This polyhedron has exactly the same number of vertices, edges and faces as in the previous case. The only difference that the isogonal decagon in the polyhedron $O[\sigma(a_1a_2a_3)]$ now becomes a regular decagon. The faces of the dual of this polyhedron is also scalene triangle. The values of the parameters $\lambda$ and $\eta$ can be obtained from (40) by letting $a_1 = a_2$.

**D2. The polyhedron** $O[\sigma(a_1a_2a_1)], a_1 \neq a_2 \neq 0$

The faces of the polyhedron consist of squares of edge length $\sqrt{2}a_1$ and isogonal decagons with isogonal hexagons of edge lengths $\sqrt{2}a_1$ and $\sqrt{2}a_2$. The number of vertices, edges and the faces are the same as those of the general orbit $O[\sigma(a_1a_2a_3)]$. Faces of the dual polyhedron is again a scalene triangle where the parameters $\lambda$ and $\eta$ can be obtained from (40) by letting $a_1 = a_2$.

**D3. The polyhedron** $O(a_1a_2a_2), a_1 \neq a_2$

The 62 faces of this polyhedron consist of the isogonal decagons, rectangles and the regular hexagons. The dual polyhedron is covered by the 62 scalene triangles whose parameters $\lambda$ and $\eta$ are determined from (40) by letting $a_3 = a_2$.

**VI. Concluding Remarks**

In this work we presented a systematic construction of the quasi regular polyhedra and their dual solids. The work is the generalization of two papers [6-7] and extensions of the Platonic-Archimedean solids, the Catalan solids to the quasi regular polyhedra and their dual solids with the use of Coxeter diagrams $A_3, B_3, H_3$.

We have employed the quaternionic simple roots to obtain the representations of the Coxeter-Weyl groups and the orbits in terms of quaternions. In particular, we have constructed the vertices of the $C_{60}$ molecule consistent with the experimental bond lengths. The important fact here that whether the polyhedra are regular (Platonic solids), semi regular (Archimedean solids) or quasi regular solids they all possess the Coxeter symmetries.

**References**

[1] F.A. Cotton, G. Wilkinson, C.A. Murillo, M. Bochmann, Advanced Inorganic Chemistry, 6[th] Ed., Wiley-Interscience, New York (1999).

[2] D.L.D Caspar and A. Klug, Cold Spring Harbor Symp. Quant. Biol. 27, 1 (1962); R. Twarock, Phil. Trans. R. Soc. A364, 3357 (2006).

[3] M. V. Jaric(Ed), Introduction to the Mathematics of Quasicrystals, Academic Press, New York (1989).

[4] H. S. M. Coxeter and W. O. J. Moser, Generators and Relations for Discrete Groups, Springer Verlag (1965).

[5] M. Koca, R.Koc, M.Al-Barwani, J. M. Phys. 44, 03123 (2003); M. Koca, R. Koc, M. Al-Barwani, J. M. Phys 47, 043507-1 (2006). M. Koca, R. Koc, M. Al-Barwani and S. Al-Farsi, Linear Alg. Appl. 412, 441 (2006). See also ref [8].

[6] M. Koca, R. Koc and M. Al-Ajmi, J. Math. Phys. 48, 113514 (2007).

[7] M. Koca, N. O. Koca and R. Koc, J. Math. Phys. 51, 043501 (2007).

[8] M. Koca, N. O. Koca and R. Koc, Quasi Regular Polyhedra and Their Duals with Coxeter Symmetries Represented by Quaternions I   arXiv:1006.2434 (submitted for publication).

[9] W. I. F. David, R. M. Ibberson, J. C. Matthewman, K. Prassides, T. J. S. Dennis, J. P. Hare, H. W. Kroto, R. Taylor, D. R. M. Walton, Nature, 353, 147 ( 1991).

[10] R.W.Carter, Simple Groups of Lie Type, John Wiley & Sons Ltd, 1972; J.E. Humphreys, Reflection Groups and Coxeter Groups, Cambridge University Press, Cambridge, 1990.

[11] M. Koca, R. Koc, M. Al-Barwani, J. Phys. A: Math. Gen. 34, 11201 (2001).

[12] R. Slansky, Phys. Rep.79, 1 (1981).

[13] M. Koca, N.O. Koca and Muna Al-Shueili, "Chiral Polyhedra Derived from Coxeter Diagrams and Quaternions" (under preparations)

[14] J.M.Hawkins, Acc. Chem. Res. 25, 150 (1992).